\documentstyle[multicol,12pt,psfig,epsfig,rotate]{article}
\begin{document}
\setcounter{page}{1}
\renewcommand{\thesection}{\Roman{section}}
\begin{center}
{\Large\bf Jamming of directed traffic on a square lattice}\\
\vspace{1cm}
\normalsize\it
{\bf Anjan Kumar Chandra}\\
Department of Physics, University of Calcutta,
92 Acharya Prafulla Chandra Road, Calcutta- 700009, India.\\
\end{center}
\begin{center}
{\bf ABSTRACT}
\end{center}
Phase transition from a free-flow phase to a jammed phase is an
important feature of traffic networks. We study this transition in the case
of a simple square lattice network for different values of data posting rate
$(\rho)$ by introducing a parameter $p$ which selects a neighbour for 
onward data
transfer depending on queued traffic. For every $\rho$ there is a critical 
value of $p$ above which
the system become jammed. The $\rho-p$ phase diagram shows some interesting
features. We also show that the average load diverges logarithmically as $p$
approaches $p_c$
and the queue length distribution exhibits exponential and algebraic nature in 
different regions of the phase diagram. 
 \\[1cm]
PACS number(s): 89.75.Hc, 89.40.-a, 89.75.Fb\\
                                                                                      
Key words: Internet traffic, free-flow phase, jammed traffic.\\

\bigskip

\section{Introduction}

The mechanism of transportation of information is important in different 
branches of science. The study of highway traffic has
been a field of interest for a long time. In view of modern perspectives,
the study of internet traffic network has also opened a new field of research.
The most important feature of these two networks is a
phase transition from a free-flow phase to a conjested phase. The main target
in constructing these networks lies in maximising the flow of data and avoiding
traffic jam. An extensive
study on internet traffic \cite{int1,int2,int3,int4} has already been done.
The study of topological structure of internet traffic network 
\cite{int12,int15} has been very helpful in increasing the efficiency of 
the network. The optimization of traffic flow has also been studied in the 
recent past \cite{int6,int8,int9}. Traffic flow has also been studied on 
scale-free network \cite{Tadic}.
In contrast to such elaborate studies on information transport in realistic 
models of complex networks (for review, see \cite{Latora}), a recent study on 
square
lattice by Mukherjee and Manna \cite{Manna} has also proved to be quite 
interesting.
Although a square lattice is a rather simplified and somewhat unrealistic
model of real-life networks, it can nevertheless capture the essential
features of traffic flow, such as the occurrence of free flow and the onset
of jamming. 

The model of Mukherjee and Manna \cite{Manna} considers an $L \times L$ square 
lattice
and a preferential direction of data flow in $-y$ direction. In a single time
step $\rho$$L$ data packets are posted randomly on the sites chosen on the
uppermost row. A single data packet is transferred to the lower left $(LL)$ 
or lower right $(LR)$ neighbour randomly from every site of the row and this
process of transferring data continues sequentially for all the rows. Finally, 
as the data packets 
reach the lowermost row (considered as sink) they are removed from the system.
The  average number of data
packets per site, $\bar{q}(\rho,t)$ was studied as a function of time for 
different values of
$\rho$. It was found that for small values of $\rho$ the steady state value 
and 
fluctuation of $\bar{q}(\rho,t)$ is small (the `free-flow' phase) and as 
$\rho$ is increased, the steady state value of $\bar{q}(\rho,t)$ increases and 
above a critical value
$\rho_{c}$, the value of $\bar{q}(\rho,t)$ increases monotonically with time 
(the `jammed' phase). 

        The most important feature in the network considered by Mukherjee and 
Manna \cite{Manna} was that, as the value of $\rho$ approaches some critical 
value $\rho_c$, the steady state value
of $\bar{q}(\rho,t)$ increases rapidly, slowing down the process of data
flow. It is due to random selection of one of the neighbours to transfer
data packets from each node. In order to investigate this feature in greater
detail, we introduce
another parameter $p$, which is the probability by which the more populated 
node is preferred to a less
populated node for onward data transfer. A choice of $p=0.5$ means random 
selection between $LL$ and $LR$ site, independent of the population of them, 
as was done in \cite{Manna}.
As $p$ is gradually decreased from $0.5$, we force the data packets
to move to the less populated neighbour and thus prevent a particular
node from being piled up by data. By decreasing the value of $p$,
the steady state value of $\bar{q}(\rho,t)$ can be decreased significantly 
even for high values
of $\rho$ below $\rho_c$, thus increasing the rate of flow through the network.
On the other hand, as $p$ is gradually increased the $\bar{q}(\rho,t)$
shows a logarithmic divergence near a critical probability  $p_c$
above which the system becomes jammed even for $\rho<\rho_c$. 
We also observe a lower limit $\rho=\rho_0$, below which the network remains in 
the free-flow phase, whatever be the value of $p$.       

 We have simulated the traffic flow for various combinations of values of 
 $\rho$ and
 $p$ for three system sizes, $L =  64, 128, 256$. In section II we present 
 the details of the model and in section III the results.    

\section{The Model}

 The network we have considered is the same as that taken by Mukherjee and
 Manna \cite{Manna}, together with a newly introduced parameter $p$ as 
described in section I. 
 Thus, on a square lattice of size $L\times L$ placed on the $x-y$ plane,
 the lattice sites are the nodes and the lattice bonds are the links.
 The data flows only along $-y$ direction. Every node can transfer data to
 any one of the two neighbours,
 one at the lower left ($LL$) and the other at the lower right ($LR$)
 position.

Each iteration in our simulation comprises of the following three
steps :
 (i) Initially $\rho$$L$ data packets are posted on the 
randomly chosen nodes of the uppermost row at $y=L$.
 (ii) The $L^2$ nodes are updated in helical sequence.
 Each node transfers a single data packet to the
      more populated neighbour with probability $p$ and to the less populated
neighbour with probability $(1-p)$. If $LL$ and $LR$ nodes are equally 
populated, any one of them is chosen randomly for transfer. 
 (iii) Finally, as a data packet reaches the bottom row at $y=0$, it 
      is removed from the system.

To characterise the flow of traffic, we measure the total number of packets
$N(t)$ in the lattice and calculate therefrom the average number of data
packets per site $\bar{q}(\rho,t) \equiv N(t)/L^2 $ (often called the mean
load) at an iteration $t$. In
the so-called {free-flow} phase, the mean load gets saturated after some
time ($10^2$ to $10^4$ iterations depending on the value of $p$ and $\rho$) 
indicating that the average fluxes of the inflow and outflow currents
of data packets have reached a steady value and are equal. In the {jammed}
phase the inflow exceeds outflow and the data accumulates continuously in the
system, resulting in a monotonically increasing mean load.  We have also
measured the number of data packets $(l)$ that are in queue at different sites
over the lattice (called the queue length) and computed its distribution 
function $q(l)$.

We have simulated upto $10^5$ iterations and
averaged over 10 configurations for $L=64$. For $L=128$ both the iteration
and number of configurations was doubled. But for low values of $\rho$, near 
$p = p_c$, the quality of data was poor and we had to iterate upto 
$3\times10^5$ time steps and average over $30$ configurations. Some runs were
repeated also for $L = 256$.
                                                                                
\section{Results }
                                                                                
The study of Mukherjee and Manna [7] was confined to $p=0.5$ and they found
that for a posting rate lower than a critical value $\rho_c=1$, the
system is in a free flow state, while for $\rho > \rho_c$ the system is in
a conjested (jammed) state. After introducing the probability $p$ for the
selection of recepient node, we observe that the above mentioned result 
(including the
precise value of $\rho_c$) remains the same for values of $p<0.5$. Thus, even
when we take care to transfer the data {\em always} to the less populated
node, the sytem does get jammed at $\rho > 1$. But as $p$ is increased above
$0.5$ the value of $\rho_c$ decreases with increase of $p$. 
                                 
The effect of introducing $p$ is most significant for high values of $\rho$ 
($0.9 < \rho < 1$). At a 
posting rate $\rho = 0.98$, the system does get jammed for $p>0.5$. For values 
of $p$ just below $0.5$ the steady state value of $\bar{q}(\rho,t)$ becomes
very high and also fluctuates highly. But if we decrease the value of $p$,
the average load per site decreases (Fig. 1) and also fluctuates much less.
Thus by controlling the parameter $p$ we are able to prevent the sites of
the lattice from being overloaded, which is very much desired in a traffic
network.
This trend remains as we decrease $\rho$ further, until the average load per 
site becomes small for any value of $p$, and there is no need to control $p$ 
anymore.

In general for a particular posting rate lower than the critical one 
($\rho < \rho_c$) the system will be in a free-flow phase upto a certain 
critical value of $p$ (say $p_c$). If we increase the value of $p$ further 
(select deliberately the more populated node more often) the system becomes 
jammed.  
For different values of $\rho$, we have determined the value $p_c$ which is  
obviously a measure of tolerance of the system
for `wrong' selection of node (Fig. 2). The solid line in the figure 
corresponds to the value of $\rho_c$ for a certain $p$ and similarly the value
of $p_c$ for a certain $\rho$.
Just below $\rho=1$ the value of $p_c$ is 1/2, but as one 
decreases
$\rho$, the system needs a larger $p$ to become jammed and the value of
$p_c$ increases. This trend continues until at some $\rho=\rho_0$ the
value of $p_c$ becomes 1 and cannot increase any more. This value of $\rho$
(which we call lower critical posting rate) is hence the minimum value of
$\rho$ below which the system always remains in the free-flow phase even 
if $p$ is set 1 i.e., the data is deliberately transferred every time to the 
more populated neighbour. It may be noted that (i) the numerical value
of $\rho_0$ depends on the system size to some extent, it is 0.24 for $L=64$
and 0.21 for $L=128$; (ii) the critical probability $p_c$
varies linearly with $\log(\rho-\rho_0)$ (Fig. 3) particularly for $L=128$.
 For higher values of $\rho$, the value of $p_c$ remains same for both system sizes, but for smaller $\rho$ the value of $p_c$ is somewhat smaller for larger 
size.
                                                                                
As mentioned above, at any posting rate in the region 
$\rho_0 < \rho < \rho_c$,
the system gets jammed and the mean load $\bar{q}(\rho,t=\infty)$ diverges
for $p > p_c$. We have observed that for high ($\rho > 0.8 $) values of $\rho$ 
this divergence is logarithmic (Fig. 4),
\begin{equation}
  \bar{q}(\rho,t=\infty) = a \log(p_c-p) + b 
\end{equation}
but for $\rho < 0.8 $ it is not logarithmic (Fig. 5). For high values of $\rho$
the curves show size dependence upto $L=128$. But for $\rho < 0.8 $ the curves
become more and more flat with increasing system size. The mean load obeys a 
scaling form with $p_c-p$ and $L$ of the following form (Fig. 5),
\begin{equation}
  \bar{q}(\rho,t)L^{0.13} = \alpha \exp[- \beta(p_c-p)]
\end{equation} 
The value of $\alpha$ is 1.85 and $\beta$ is 3.2.

       Now just as $p_c$ varies with $\rho$, the quantity $\rho_c$ also varies 
with $p$ (Fig. 2). We have
studied the behaviour of the mean load as a
function of $\rho$ for a given value of $p$ (Fig. 6). It is clear
from the figure that as the value of $\rho$ decreases the mean load becomes
very small irrespective of the value of $p$.
Moreover, for a given $p$, when $\rho$ is less than 
$\rho_c$, the 
system is in a free-flow phase and the rate of flow increases with an increase 
of posting rate. However, as the posting rate exceeds $\rho_c$, the rate of flow
drops suddenly and the system is jammed. This effect is called
{\em hysteresis} in the context of traffic flow \cite{Chowdhury}.
For all values of $p$ the mean load diverges algebraically with $\rho_c-\rho$,
\begin{equation}
  N(\rho,t)/L^2 \propto (\rho_c-\rho)^{-\lambda}
\end{equation}  
The exponent $\lambda$ is $1.0$ for $p=0.5$ as observed in \cite{Manna} but
decreases for $p$ above and below $0.5$ (Fig. 7). Thus it decreases to $0.49$ 
at $p=0.4$ and then remains constant even if $p$ is decreased further. Again 
if $p$ is increased above $0.5$ the exponent decreases very rapidly to $0.28$ 
at $p=0.6$.

               The fluctuation of mean load as a function of $\rho_c-\rho$ for
different values of $p$ exhibits almost a similar pattern like the mean load.
The fluctuation is maximum for $p=0.5$. As $p$ is gradually increased or 
decreased from $0.5$ the fluctuation decreases. It is probably due to 
the fact that when $p=0.5$ we are selecting the recipient node randomly, 
whereas in other cases we are preferring a certain type of node (more populated
or less populated) each time we forward a data.

 To gain further insight into the change of the
behaviour of the mean load as a function of $p$ and $\rho$, we have
studied the {\em queue length distribution} function $q(l)$ which is the 
distribution of the number of data packets $(l)$ present at different sites. 
At any posting rate
below $\rho_c$, as the system is always jammed for $p > p_c$, we shall confine 
our study of queue length distribution only in the {free-flow phase},
i.e. for $p < p_c$.
At $p=0.5$ the queue length distribution is exponential for all values of 
$\rho$ in accordance with the result of Mukherjee and Manna. According to our
study the queue length distribution is exponential for any value of $p < p_c$ 
above a certain value of $\rho$ (dependent on system size). But below this
value of $\rho$, the queue length distribution is algebraic (Fig. 8) only for 
a very narrow range of value of $p$ (shaded region in Fig. 2). Below this 
narrow range of value of $p$ the queue length distribution is exponential (Fig. 9).          

\begin{figure}
\noindent \includegraphics[clip,width= 6cm, angle = 270]{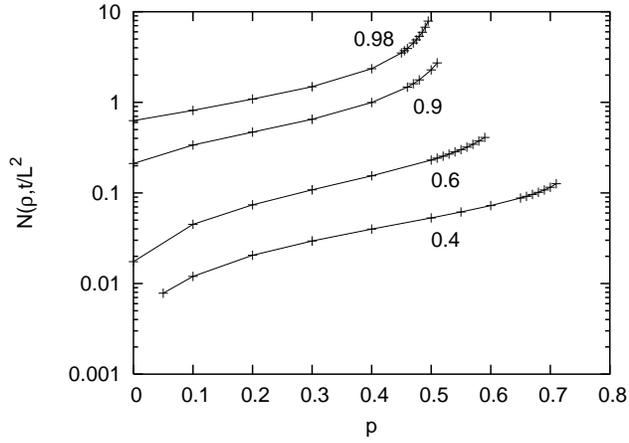}
\caption{ \small { Plot of mean load $N(t)/L^2$ as a function of $p$ for 
different values of $\rho$.} }
\end{figure}

\begin{figure}
\noindent \includegraphics[width= 12cm, angle = 0]{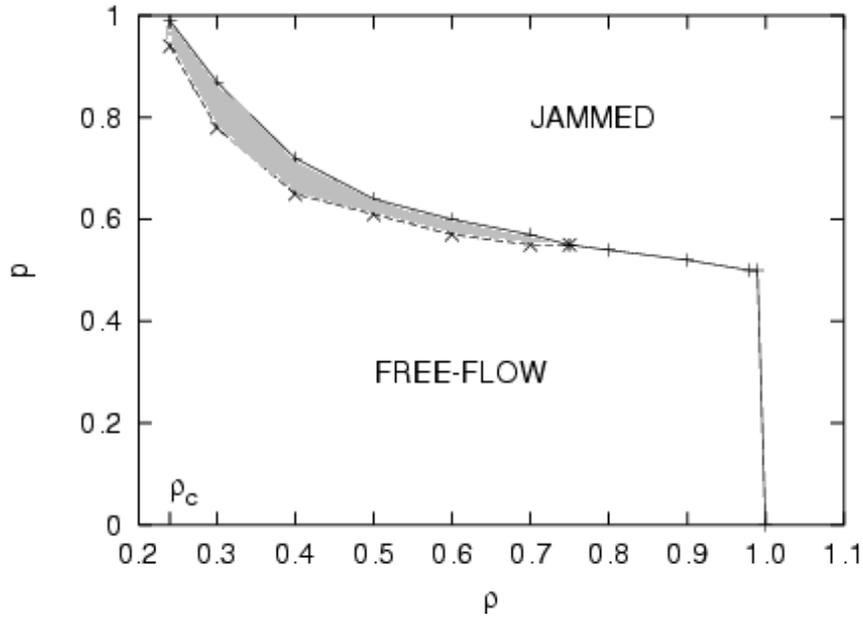}
\caption{ \small {Various phases of the network for different
values of $\rho$ and $p$. The solid line is the boundary between the jammed and
the free-flow phase and also the locus of ($\rho_c$, $p_c$). The broken line separates the regions of algebraic and
exponential queue-length distribution (both) within the free-flow phase. The
algebraic nature of $q(l)$ exists only in the shaded region.}}
 \end{figure}

\begin{figure}
\noindent \includegraphics[clip,width= 6cm, angle = 270]{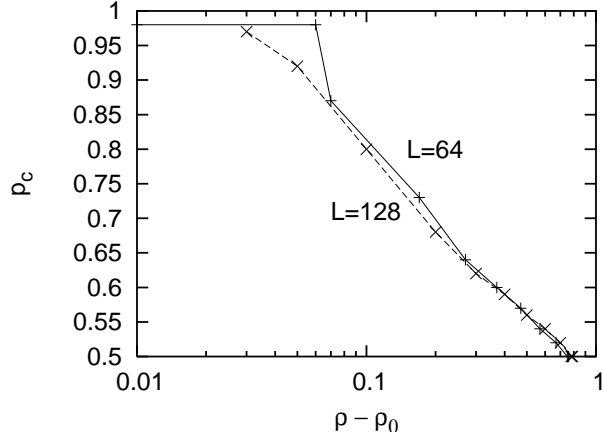}
\caption{ \small { Plot of critical value $p_c$ as a function of $\rho - \rho_0$
 for system sizes $L=64$ and $L=128$. The curves fit roughly to 
$-0.155\log(\rho - \rho_0) + 0.45$}} 
\end{figure}

\begin{figure}
\noindent \includegraphics[clip,width= 6cm, angle = 270]{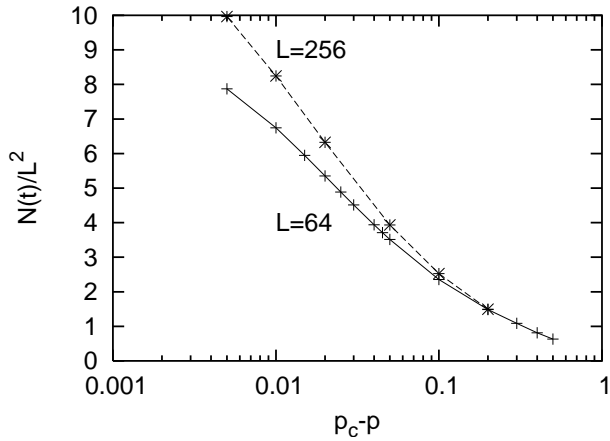}
\caption{ \small {Plot of average load per site $N(t)/L^2$ as a function of 
$p_{c}-p$ for $\rho=0.98$ for system sizes $L$ = 64, 128 and 256. The linear 
portion of the curves fit to $-1.9\log(p_c-p)-2.16$ for $L=64$ and to 
$-2.58\log(p_c-p)-3.75$ for $L=256$. The curve for $L=128$ coincides with
$L=256$} }
\end{figure}

\begin{figure}
\noindent \includegraphics[clip,width= 6cm, angle = 270]{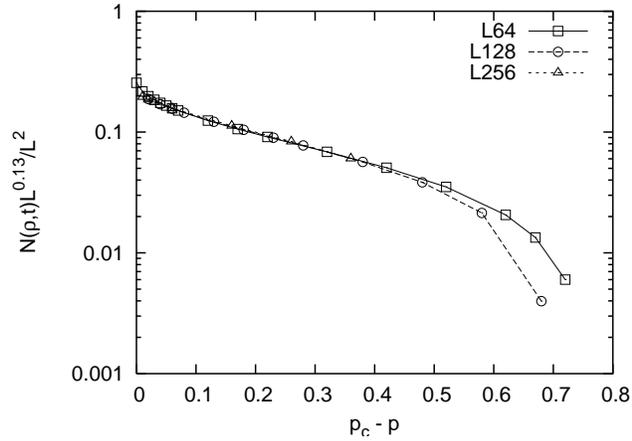}
\caption{ \small {Scaled plot of the average load per site $N(t)/L^2$ as a 
function of $p_{c}-p$ for $\rho=0.40$ and system sizes $L$ = 64, 128 and 256.
The scaled plot fits nicely in a normal-log graph. }}
\end{figure}

\begin{figure}
\noindent \includegraphics[clip,width= 6cm, angle = 270]{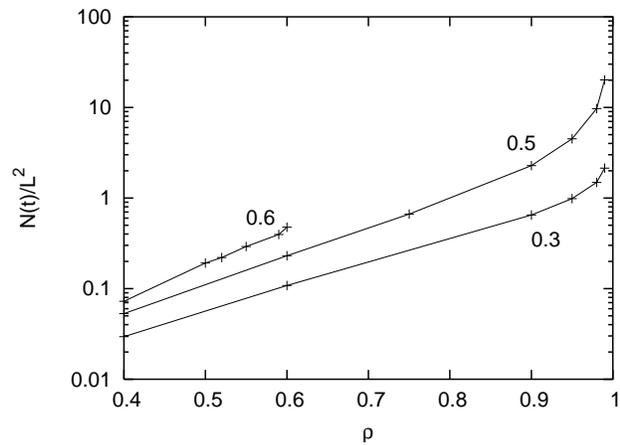}
\caption{ \small { Plot of mean load $N(t)/L^2$ as a function of $\rho$ for
different values of $p$. The numbers labelling the curves correspond to the 
various values of $p$.} }
\end{figure}

\begin{figure}
\noindent \includegraphics[clip,width= 6cm, angle = 270]{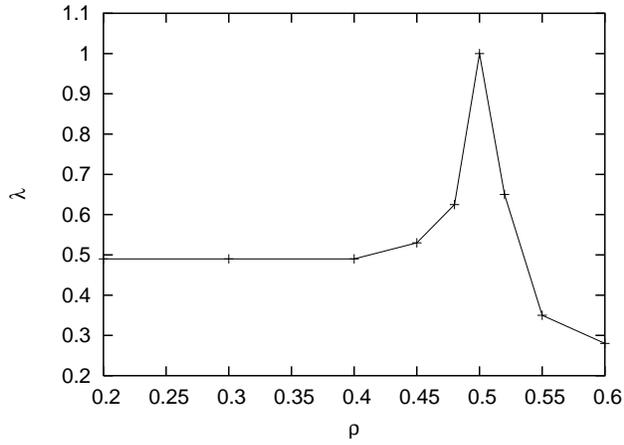}
\caption{ \small {Exponent ($\lambda$) of mean load for different values of 
$\rho$ (see Eq. 3). The exponent is maximum for $\rho=0.5$. }}
\end{figure}

\begin{figure}
\noindent \includegraphics[clip,width= 8cm, angle = 270]{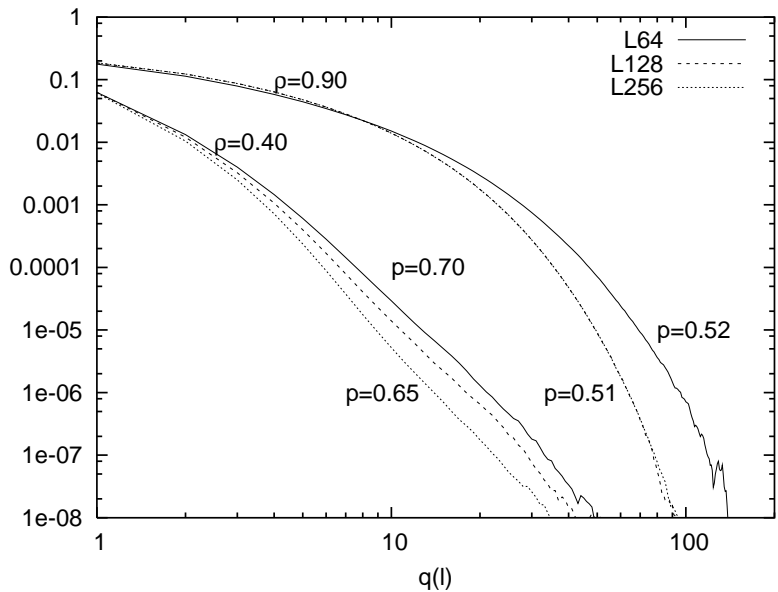}
\caption{ \small {Queue length distribution for $\rho=0.40$ and 
$\rho=0.90$ for $L=$ 64, 128 and 256 in log-log scale. The values of 
corresponding $p$ values are shown beside the plots.
 For $\rho=0.40$ the distribution is algebraic near $p=p_c$. }}
 \end{figure}      

\begin{figure}
\noindent \includegraphics[clip,width= 8cm, angle = 270]{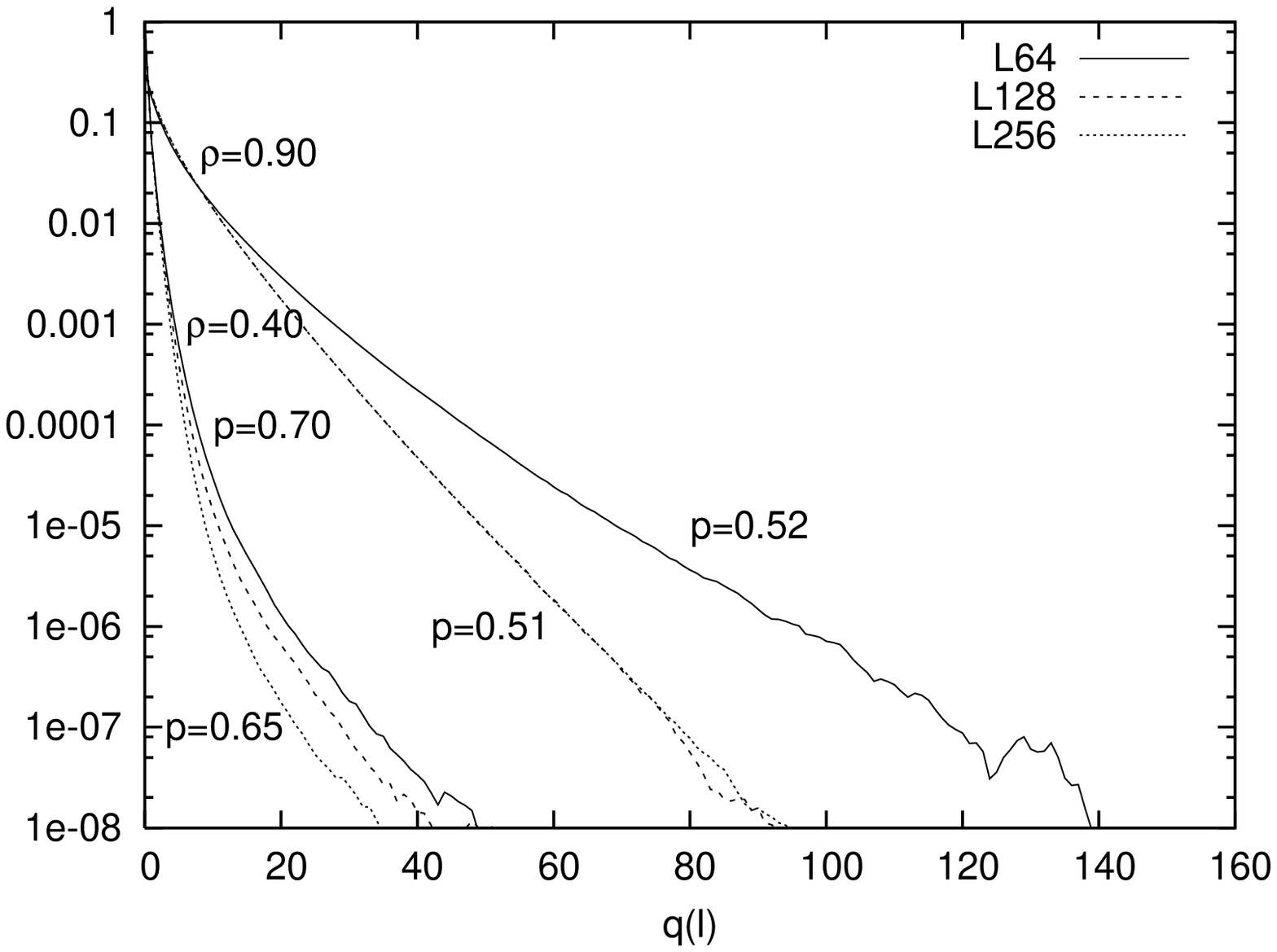}
\caption{ \small {Queue length distribution for $\rho=0.40$ and
$\rho=0.90$ for $L=$ 64, 128 and 256 in normal-log scale. For
$\rho=0.90$ the distribution is exponential for any value of $p$ less than
$p_c$. }}
 \end{figure}

In summary, we have studied the effect of introducing a probability $p$ for 
choosing the more populated node (for onward data transfer) in a traffic 
network. For a posting rate below a lower critical value
$\rho_0$, the system is in a free-flow phase for any $p$, but for 
$\rho > \rho_0$ the system undergoes a transition from a free-flow phase to a 
jammed
phase as $p$ exceeds a critical value $p_c (< 1)$. As $\rho$ increases from
$\rho_0$ to the critical value $\rho_c$, the value of $p_c$ also decreases from
1 to 0.

{\bf Acknowledgement :} The author is grateful to S. Dasgupta, S.S. Manna,
G. Mukherjee and P.Sen for encouragement and fruitful discussions. The work was
 supported by UGC fellowship.

\end{document}